\documentclass[prb,showpacs,amsmath,amssymb,aps,twocolumn]{revtex4}

\usepackage{graphicx}

\begin{document}
\draft
\title{Optical properties of (In,Ga)As capped InAs quantum dots grown on [11k] substrates}
\author{V. Mlinar}
\email{vladan.mlinar@ua.ac.be}
 \affiliation{Departement
Fysica,Universiteit Antwerpen, Groenenborgerlaan 171, B-2020
Antwerpen, Belgium}
\author{F. M. Peeters}
\email{francois.peeters@ua.ac.be}
\affiliation{Departement
Fysica,Universiteit Antwerpen, Groenenborgerlaan 171, B-2020
Antwerpen, Belgium}

\date{\today}

\begin{abstract}
Using three-dimensional \mbox{${\bf k}\!\cdot\!{\bf p}$} calculation
including strain and piezoelectricity, we showed that the size of
the quantum dot (QD) in the growth direction determines the
influence of the (In,Ga)As capping layer on the optical properties
of [11k] grown InAs QDs, where k=1,2,3. For flat dots, increase of
$In$ concentration in the capping layer leads to a decrease of the
transition energy, as is the case of [001] grown QDs, whereas for
large dots an increase of the $In$ concentration in the capping
layer is followed  by an increase of the transition energy up to a
critical concentration of $In$, after which the optical transition
energy starts to decrease.
\end{abstract}


\maketitle

Manipulation of quantum dots' (QDs) properties is driven by current
and potential applications, ranging from  QD lasers, and
photodetectors to single polarized photon sources. Growth
conditions, such as growth temperature, substrate orientation, or
capping procedures, determine the QD electronic and optical
properties.~\cite{Shchukin} In order to produce good quality QD
structures with high densities and low size dispersion, or to
control lateral and vertical ordering of QDs in QD lattices, growth
on high index surfaces has been put forward.
\cite{Schmidbauer,Caroff,Celebi,Jacobi2} Furthermore, to achieve
long wavelength emission, e.g. larger than 1.3$\mu$m in a QD laser
diode, or alternatively user-defined detection wavelength, e.g. for
quantum dot infrared photodetectors, QD-in-a-well (DWELL) structures
were introduced. \cite{Lin,Gong} Namely, optical properties of a QD
are tuned by size and chemical composition of QW layer, where the QD
is embedded in.

In a widely investigated DWELL system, [001] grown InAs QD embedded
in In$_x$Ga$_{1-x}$As QW, variation of the $In$ concentration in the
capping layer as well as the thickness of the layer influence the
hydrostatic component of the strain tensor and consequently the
transition energies: \uppercase{i}ncrease of the $In$ concentration
in the QW leads to a decrease of the transition energy. What is
happening in the case of QDs grown on [11k] substrates, where
k=1,2,3? How does the (In,Ga)As capping layer influence the optical
properties of the InAs QDs grown on [11k] substrates? In this Letter
we answer these questions and provide a guideline for the variation
of the transition energy of [11k] grown QDs, as function of the
capping layers thickness and chemical profile, and for different dot
composition.

Prior to understanding how the capping layer influences the
transition energies of [11k] grown InAs QDs, one has to know the
effect on the transition energies of QD growth on [11k] substrates
(k=1,2,3). The origin of the variation of the transition energy with
the substrate orientation can be traced back to the competition of
several effects: \cite{MlinarAPL} (i) hydrostatic component of the
strain tensor is responsible for a shift of the conduction band
upwards and the valence bands downwards, (ii) biaxial component of
the strain tensor influences the degree of the valence band mixing,
and (iii) variation of the hole effective mass with the substrate
orientation, which can significantly alter the effects of the size
quantization in the QD. Actually, we have shown that the QD size in
the growth direction determine which of the three above mentioned
effects will be the dominant one, regardless on the dot shape.
Therefore, we consider here two model lens-shaped QDs with different
height: L1 QD with radius R=9.04nm and height h=4.52nm, and L2 QD
with radius R=9.04nm and height h=9.04nm. The thickness of the
capping layer is assumed to be the same as the height of the dots,
whereas the $In$ concentration in the capping layer is varied from 0
to 30$\%$.

\begin{figure}[tpb]
  \begin{center} \includegraphics[width=0.9\linewidth]{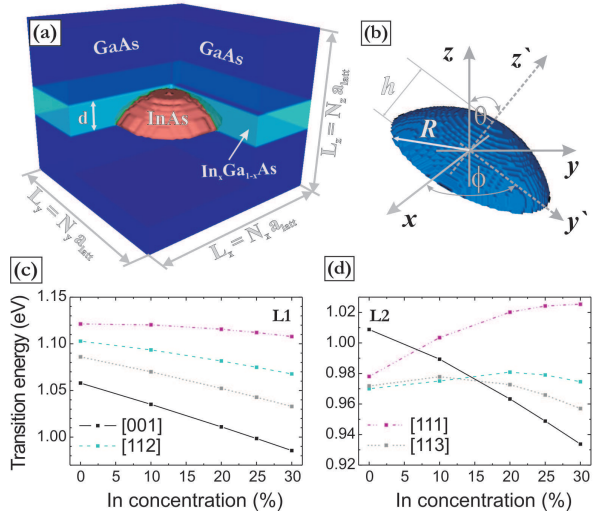}
    \caption{(color online) (a) Model InAs QD. (b) Transformation of the general [11k] coordinate system
    to the conventional [001] coordinate system. Transition
energies of L1 (c) and L2 (d) QDs as they vary with the $In$
concentration in the capping layer for different substrate
orientations.}
    \label{fig:first}
  \end{center}
\end{figure}

A model QD, as it enters our calculations, is constructed on a
three-dimensional (3D) rectangular grid with a grid step equal to
the lattice constant of GaAs, and is shown in Fig.
\ref{fig:first}(a). In our full 3D model, the strain distribution is
calculated using continuum elasticity and the single particle states
are obtained from an eight-band \mbox{${\bf k}\!\cdot\!{\bf p}$ }
theory~\cite{Mlinar1} including strain and piezoelectricity. In
order to properly take into account the effect of the different
substrate orientation, the coordinate system is rotated in a way
that the Cartesian coordinate z$'$ coincides with the growth
direction [Figs.~\ref{fig:first}(b)].~\cite{Henderson} The general
[11k] coordinate system $(x',y',z')$ is related to the conventional
[001] system $(x, y, z)$ through a transformation matrix
U=U$(\phi,\theta)$. The angles $\phi$ and $\theta$ represent the
azimuthal and polar angles, respectively, of the [11k] direction
relative to the [001] coordinate system. Transition energies are
calculated taking into account the direct Coulomb interaction.

What will happen when both effects, QD growth on high index surfaces
and capping, are present? Transition energies of L1 and L2 QD,
extracted from our numerical calculations, as they vary with
substrate orientation and $In$ concentration in the capping layer,
are shown in Figs. \ref{fig:first}(c) and (d), respectively. On can
see that for both, [001] grown L1 and [001] grown L2 QD, an increase
of the $In$ concentration in the capping layer leads to a decrease
of the transition energy. On the other hand, our findings on the
transition energies versus $In$ concentration for [11k] grown QD,
depend heavily on the dot size in the growth direction. Let us first
discuss the case of the L1 QD.

\begin{figure}[tpb]
  \begin{center} \includegraphics[width=0.75\linewidth]{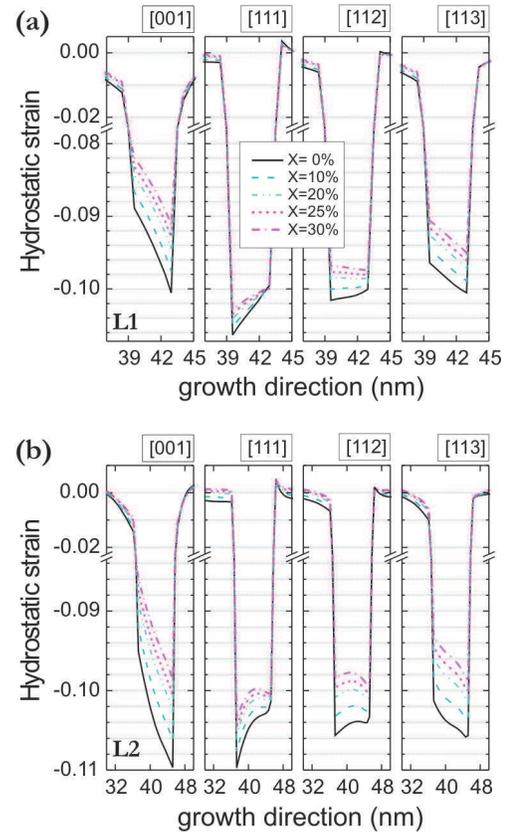}
    \caption{(color online) Hydrostatic component of the strain tensor as it varies with the substrate orientation
    and $In$ concentration in the capping layer of L1 QD (a) and L2 QD (b).}
    \label{fig:second}
  \end{center}
\end{figure}

The variation of the transition energy with the $In$ concentration
does not depend qualitatively on the substrate orientation, i.e.
with increase of $In$ concentration the transition energy decreases,
as was the case for [001] grown QDs [black line in Fig.
\ref{fig:first}(a)]. This is a simple consequence of the variation
of the hydrostatic component of the strain tensor with the substrate
orientation and $In$ concentration in the capping layer, as shown in
Fig. \ref{fig:second}(a).
The hydrostatic component of the strain tensor of [11k] grown QDs
reduces with the increase of the $In$ concentration, as in the case
of [001] grown QDs. The substrate orientation only determines the
degree of the influence of the capping layer on the hydrostatic
strain, and consequently on the transition energy. Namely, for [111]
grown QDs, transition energies decrease slower with the increase of
the $In$ concentration, whereas same dependence for [113] grown QDs
is similar to the reference case of [001] grown QDs.

\begin{figure}[tpb]
  \begin{center} \includegraphics[width=0.85\linewidth]{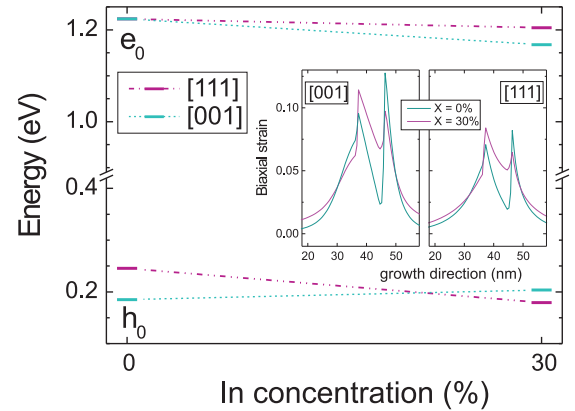}
    \caption{(color online) Variation of the electron and hole energy
    levels of [001] (short dash) and [111] (dash-dot-dot) L2 QD with $In$ concentration in the capping layer. Insets show the variation of the
biaxial component of the strain tensor with $In$ concentration of
[001] and [111] grown L2 QD.}
    \label{fig:third}
  \end{center}
\end{figure}

However, for the L2 QD surprising results are obtained: with the
increase of the $In$ concentration in the capping layer the
transition energies of [11k] grown L2 QD increase, exactly the
opposite to the dependence for [001] grown L2 QD. After the
concentration of $In$ in the capping layer reaches some critical
value, $\sim$ 35$\%$, $\sim$20$\%$, and $\sim$10$\%$, for [111],
[112], and [113] grown L2 QDs, respectively, the transition energy
starts to decrease with consecutive increase of $In$ concentration
in the capping layer [it starts to follow the pattern of [001] grown
L2 QD]. What is the origin of such a behavior? Dependence of the
hydrostatic strain on the substrate orientation and $In$
concentration in the capping layer is shown in Fig.
\ref{fig:second}(b) demonstrating that the increase of the $In$
concentration leads to a decrease of the hydrostatic strain, as in
the case of [001] grown QDs. We single out the most pronounced case,
[111] grown L2 QD, and show in Fig. \ref{fig:third} calculated
electron and hole ground state for x=0$\%$ and x=30$\%$ $In$
concentration in capping layer. As a comparison, we show the results
for [001] grown L2 QD as well. Clearly, the variation of the hole
ground state, which is most strongly influenced by the strain, with
the capping leads to an increase of the transition energy. It is
caused by the increase of the biaxial component of the strain, as
shown in the inset of Fig. \ref{fig:third}. Actually, the
competition between the increased biaxial component of the strain
tensor, responsible for the decrease of the valence band mixing, and
the decrease of the hydrostatic strain with increase of the $In$
concentration determine the transition energy. Note also that even
for [001] grown L1 QD, biaxial strain is increased, but the
hydrostatic strain has a dominant influence on the transition
energy.

\begin{figure}[tpb]
  \begin{center} \includegraphics[width=0.7\linewidth]{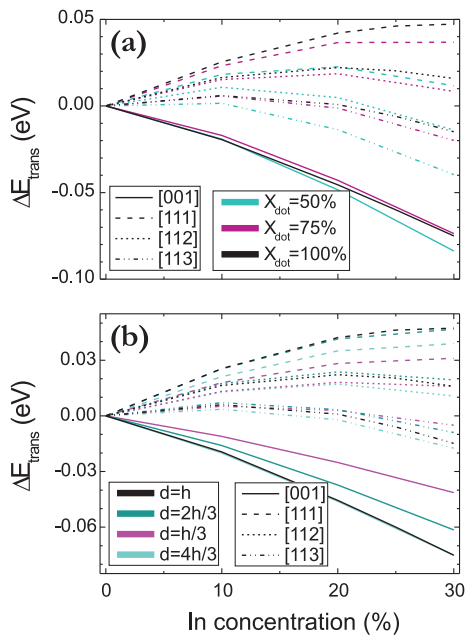}
    \caption{(color online) $\Delta$E$_{trans}$=E$_{trans}$(x)-E$_{trans}$(0), where x is $In$ concentration in the capping layer,
    of L2 QDs, versus $In$ concentration in the capping layer when (a)
    the dot composition is varied from pure InAs to
    In$_{0.5}$Ga$_{0.5}$As and In$_{0.75}$Ga$_{0.75}$As; (b) the
    thickness of the capping layer, d, is varied from d=h, where h is
    the dot height, to d=h/3, d=2h/3, and d=4h/3.}
    \label{fig:fourth}
  \end{center}
\end{figure}

At the end we address how the above conclusions are affected by the
variation in the dot composition and the thickness of the capping
layer, since those are the uncertainties to be expected in the
experiment. For that purpose we modified the L2 QD composition from
pure InAs to In$_{0.5}$Ga$_{0.5}$As and In$_{0.75}$Ga$_{0.75}$As,
and thickness of the capping layer, d, from d=h, where h is the L2
dot height, to d=h/3, d=2h/3, and d=4h/3. Our results are shown in
Figs. \ref{fig:fourth}(a) and (b). An increase of the transition
energy with increase of the $In$ concentration in the capping layer
is observed regardless of the variation of the dot composition or
capping layer thickness. Note that the critical $In$ concentration
after which the transition energy vs. $In$ concentration dependence
starts to follow the expected is reduced for L2 QD with 50$\%$Ga in
the dot, as it can be seen in Fig.~\ref{fig:fourth}(a). For example,
for [111] grown pure InAs QD critical In concentration is 35$\%$,
whereas for In$_{0.5}$Ga$_{0.5}$As QD critical $In$ concentration in
the capping layer is 20$\%$.

In conclusion, our 3D \mbox{${\bf k}\!\cdot\!{\bf p}$} calculation
including strain and piezoelectricity showed that the QD size in the
growth direction determines the influence of the (In,Ga)As capping
layer on the optical properties of [11k] grown InAs QDs, where
k=1,2,3. For flat dots, an increase of $In$ concentration in the
capping layer leads to a decrease of the transition energy, as it is
the case of [001] grown QDs, whereas the large dots exhibit an
opposite behavior i.e. increase of the transition energy with
increase of $In$ concentration up to a critical $In$ concentration
after which the transition energies start to decrease. We have shown
that our conclusions were not sensitive on the dot composition and
thickness of the capping layer, therefore possible to verify
experimentally.

This work was supported by the Belgian Science Policy (IAP), and the
European Union Network of Excellence: SANDiE.

\end{document}